\documentclass[fleqn,12pt,twoside]{article}
\usepackage[headings]{espcrc1}

\readRCS
$Id: espcrc1.tex,v 1.2 2004/02/24 11:22:11 spepping Exp $
\usepackage{graphicx}
\usepackage[figuresright]{rotating}

\newcommand{\AmS}{{\protect\the\textfont2
  A\kern-.1667em\lower.5ex\hbox{M}\kern-.125emS}}

\title{Unintegrated parton distributions 
       and pion production in $pp$ collisions
       at~SPS and RHIC energies} 
 
\author{A. Szczurek
\address[MCSD]{Institute of Nuclear Physics, \\
PL-30-059 Cracow, Poland}  
\address[MCSD]{University of Rzesz\'ow, \\
PL-35-959 Rzesz\'ow, Poland} }

\runtitle{1-column format camera-ready paper in \LaTeX}
\runauthor{A. Szczurek}

\begin{document}

\maketitle

\begin{abstract}
Inclusive cross sections for pion production
in $pp$ collisions are calculated for the first time
fully based on unintegrated gluon, quark and antiquark
distributions (uPDF).
We use recently developed Kwieci\'nski uPDF's and
different phenomenological fragmentation functions (FF)
from the literature.
In addition to the $gg \to g$ diagram we include also
$g q \to q$ and $q g \to q$ diagrams
for inclusive parton production. The cross section for pions is
obtained then by convoluting the inclusive parton distributions and
FFs. Applications for SPS and RHIC are shown.
\end{abstract}

\section{INTRODUCTION}

The distributions of mesons at large transverse momenta
in $p p$ or $p \bar p$ collisions are usually calculated
in the framework of perturbative QCD using collinear factorization
(see e.g. \cite{Owens,Field})
In order to extend the calculation towards lower values of
meson transverse momenta it was suggested to add an extra
Gaussian distribution in transverse momentum
\cite{Wang2000,Levai_LO}.
It becomes clear that this procedure is effective in
the following sense.
The transverse momentum originates either from the nonperturbative
``really internal'' momentum distributions of partons in nucleons
(of the order of a fraction of GeV) and/or is generate dynamically
as the inital state radiation process (of the order of GeV).
In principle, the second component depends on the value of
longitudinal momentum fraction.

Recently Kwieci\'nski and collaborators
\cite{GKB03} have shown how to solve the so-called
CCFM equations by introducing uPDF's in
the space conjugated to the transverse momenta.
In the following we shall use those uPDF's for pion production
at SPS, ISR and RHIC energies.

\section{FORMALISM}

\begin{figure}[!thb] 
\begin{center}
\includegraphics[width=3.0cm]{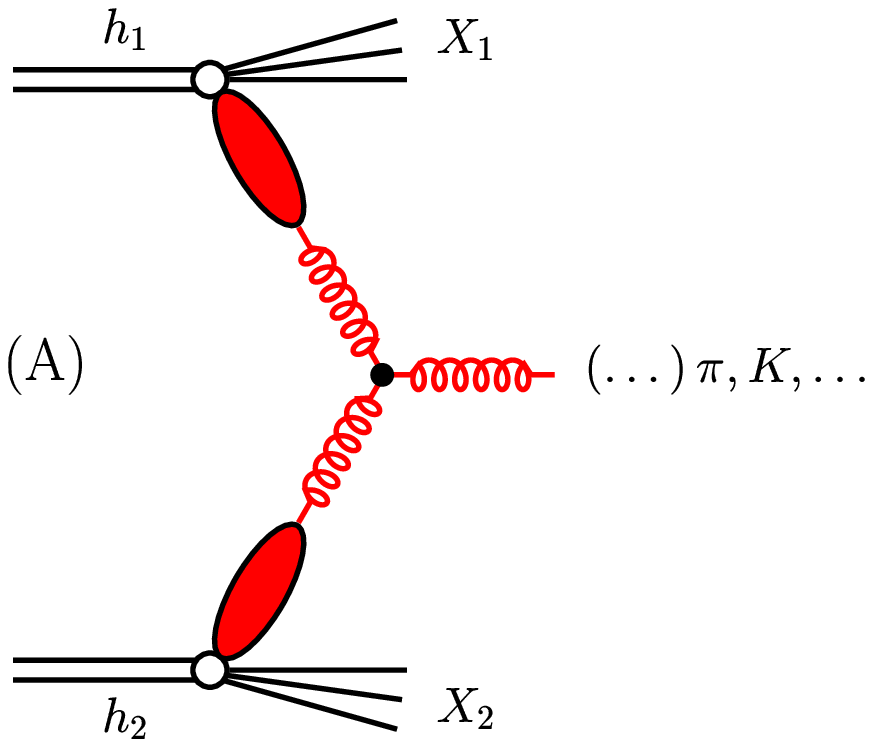}
\includegraphics[width=3.0cm]{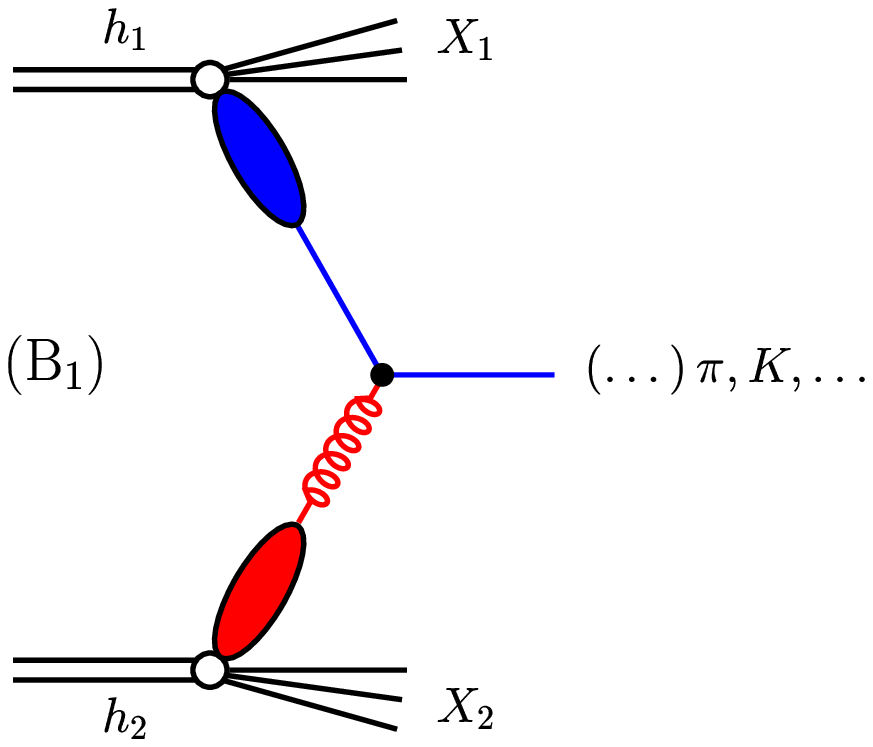}
\includegraphics[width=3.0cm]{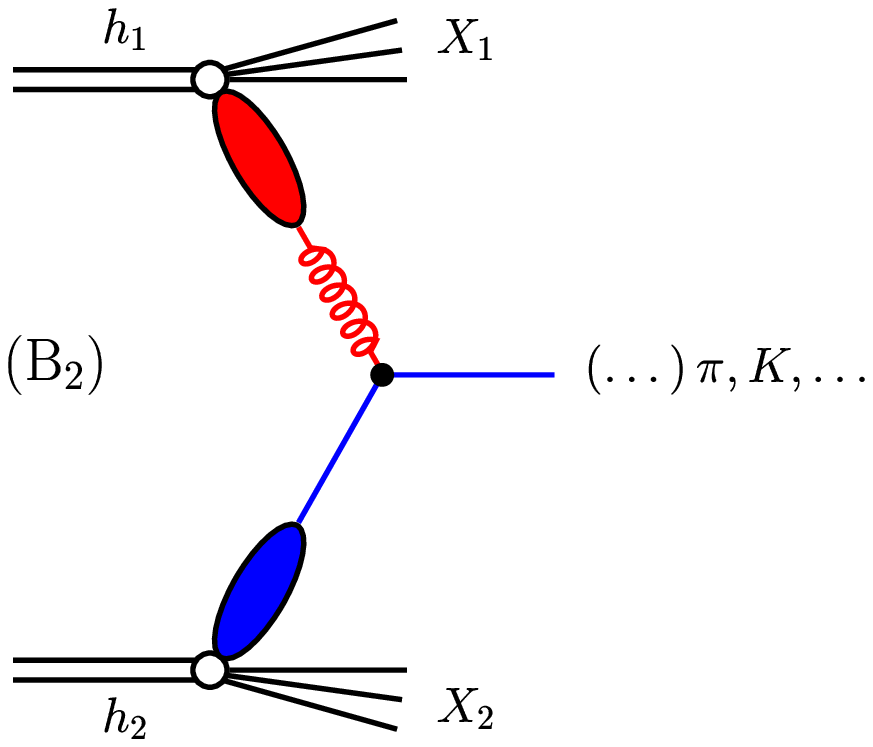}
\caption[*]{Leading-order diagrams for inclusive parton production
\label{fig:diagrams}
}
\end{center}
\end{figure}

The approach proposed by Kwieci\'nski is very convenient to
introduce the nonperturbative effects like
internal (nonperturbative) transverse momentum distributions
of partons in nucleons.
It seems reasonable, at least in the first approximation,
to include the nonperturbative effects in the factorizable way
\begin{equation}
\tilde{f}_q(x,b,\mu^2) = 
\tilde{f}_q^{pert}(x,b,\mu^2)
 \cdot F_q^{np}(b) \; .
\label{modified_uPDFs}
\end{equation}
In the following, for simplicity, we use a flavour and
$x$-independent form factor
\begin{equation}
F_g^{np}(b) = F_q^{np}(b) = F_{\bar q}^{np}(b) = 
 F^{np}(b) = \exp\left(-\frac{b^2}{4 b_0^2}\right) \; . 
\label{formfactor}
\end{equation}
Within $k_t$-factorization approach there are three basic diagrams
for inclusive parton production shown in Fig.\ref{fig:diagrams}.
The formulae for inclusive parton production,
after some minor approximation \cite{CS05},
can be written in the equivalent and compact way in the so-called
impact parameter space\\
for diagram A:
\begin{eqnarray}
&&\frac{d \sigma^{A}}{dy d^2 p_t} = \frac{16  N_c}{N_c^2 - 1}
{\frac{1}{p_t^2}}
 \alpha_s({p_t^2})
\int
{\tilde f}_{g/1}(x_1,b,\mu^2) \;
{\tilde f}_{g/2}(x_2,b,\mu^2)
 J_0(p_t b) \; 2 \pi b db    \; , 
\label{diagram_A_b}
\end{eqnarray}
for diagram B$_1$:
\begin{eqnarray}
&&\frac{d \sigma^{B_1}}{dy d^2 p_t} = \frac{16  N_c}{N_c^2 - 1}
\left( \frac{4}{9} \right)
{\frac{1}{p_t^2}}
 \alpha_s({p_t^2})
\sum_f \int
{\tilde f}_{q_f/1}(x_1,b,\mu^2) \;
{\tilde f}_{g/2}(x_2,b,\mu^2)
 J_0(p_t b) \; 2 \pi b db    \; , 
\label{diagram_B1_b}
\end{eqnarray}
for diagram B$_2$:
\begin{eqnarray}
&&\frac{d \sigma^{B_2}}{dy d^2 p_t} = \frac{16  N_c}{N_c^2 - 1}
\left( \frac{4}{9} \right)
{\frac{1}{p_t^2}}
 \alpha_s({p_t^2})
\sum_f \int
{\tilde f}_{g/1}(x_1,b,\mu^2) \;
{\tilde f}_{q_f/2}(x_2,b,\mu^2)
 J_0(p_t b) \; 2 \pi b db    \; . 
\label{diagram_B2_b}
\end{eqnarray}

The Kretzer FFs (see e.g.\cite{Kretzer2000}) are used
to convert partons to hadrons. Details of the hadronization procedure
are explained in \cite{CS05}.

\section{RESULTS}

\subsection{SPS energies}

\begin{figure}[htb] 
\begin{center}
    \includegraphics[width=5.0cm]{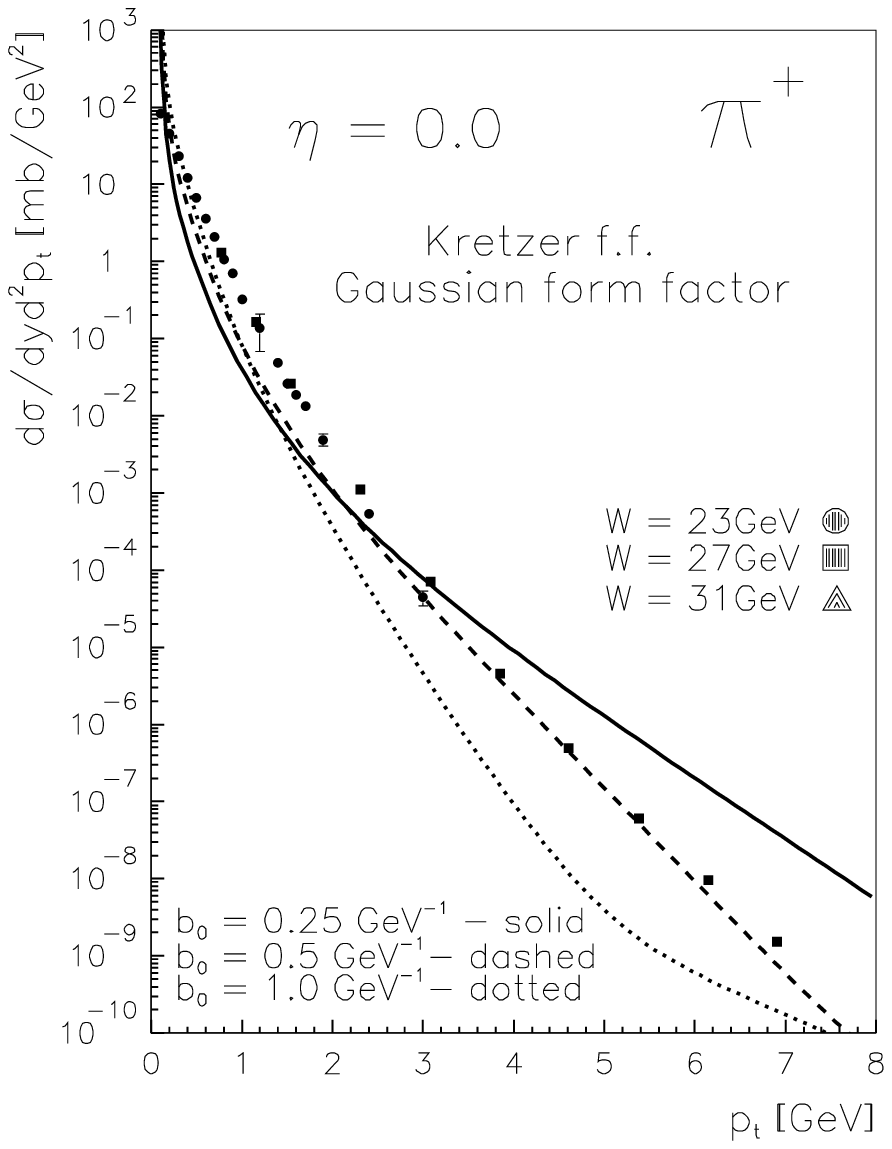}
    \includegraphics[width=5.0cm]{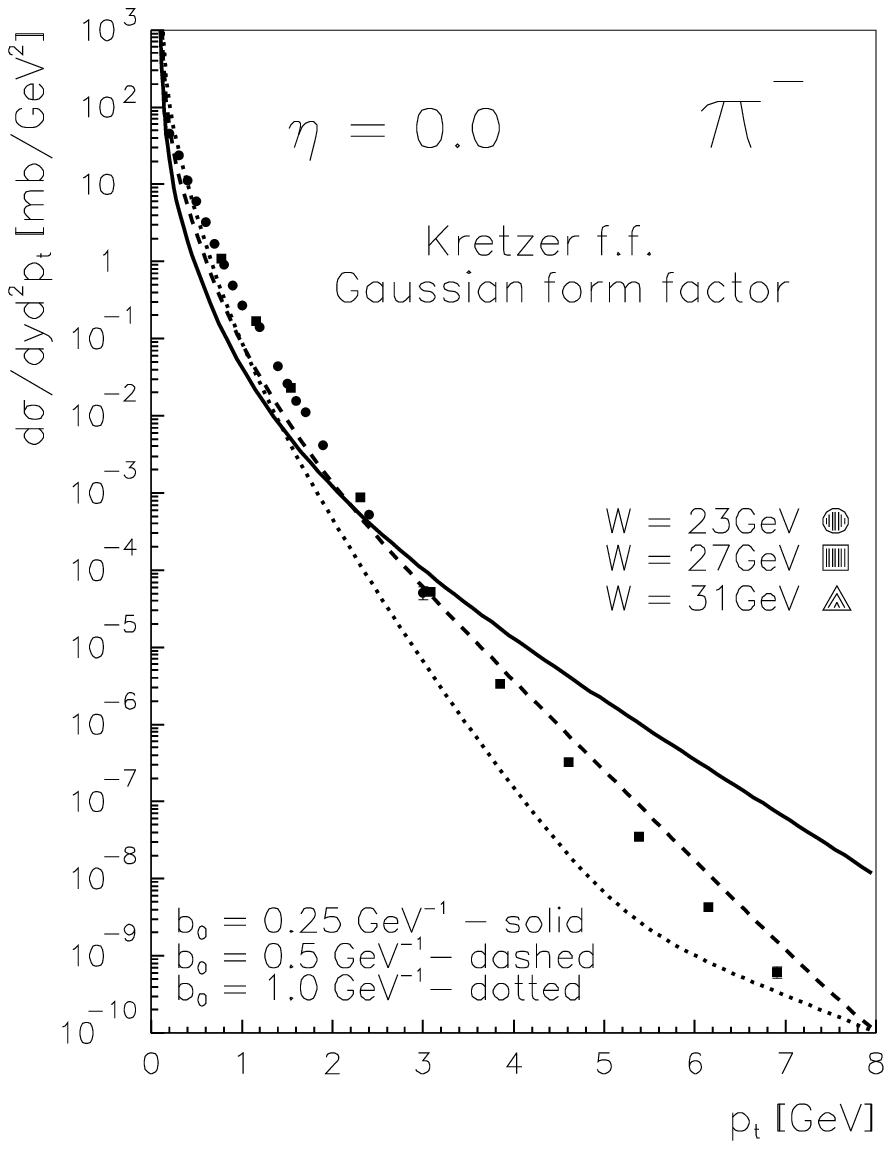}
\caption[*]{
Invariant cross section as a function of transverse momentum
of $\pi^+$ (left panel)
and $\pi^-$ (right panel) for $\eta$ = 0, W = 27.4 GeV and
different values of $b_0$.
Experimental data for W  = 23, 31 GeV from \cite{Alper} and
for W = 27.4 GeV \cite{Antreasyan} are shown for comparison.
\label{fig:pions_pt}
}
\end{center}
\end{figure}

In Fig.\ref{fig:pions_pt} we compare our model invariant cross
sections for $p p \to \pi^+$ (left panel) and $p p \to \pi^-$
(right panel) as a function of pion transverse momentum
at W = 27.4 GeV for different values of the parameter $b_0$ in
Eq.(\ref{formfactor}).
In principle, our
result should not exceed experimental data especially in
the perturbative regime of $p_t >$ 2 GeV where the perturbative
$2 \to 2$ parton subprocesses are crucial.
This limits the value of the nonperturbative
form factor to $b_0 >$ 0.5 GeV$^{-1}$.  

\subsection{RHIC energies}

\begin{figure}[!thb] 
\begin{center}
\includegraphics[width=4.5cm]{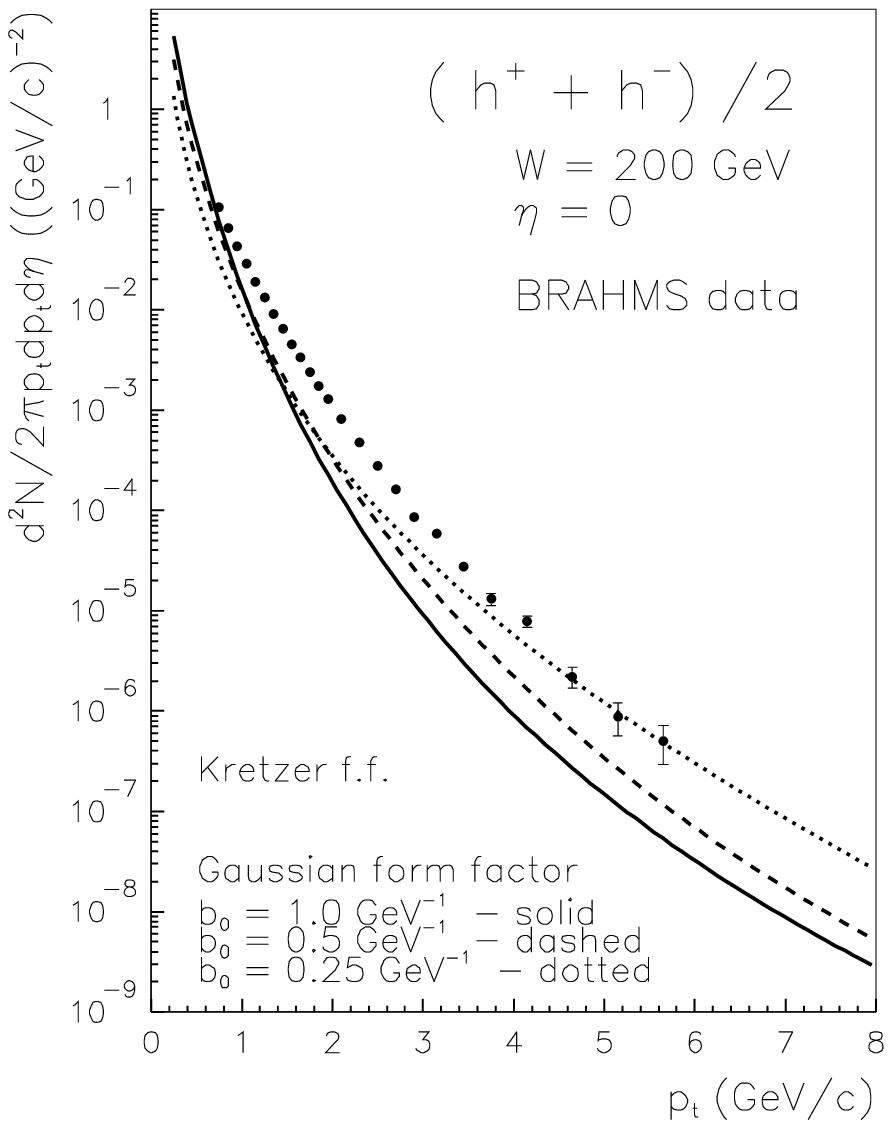}
\includegraphics[width=4.5cm]{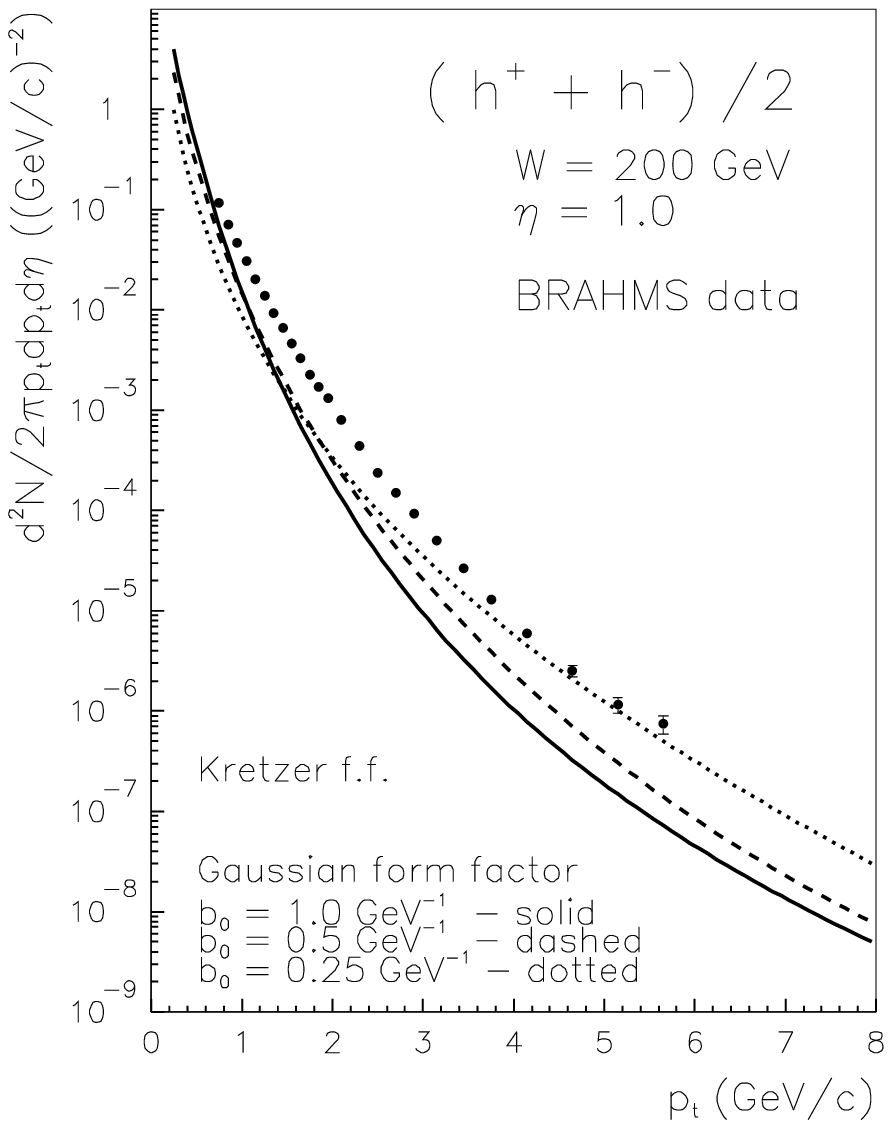}\\
\includegraphics[width=4.5cm]{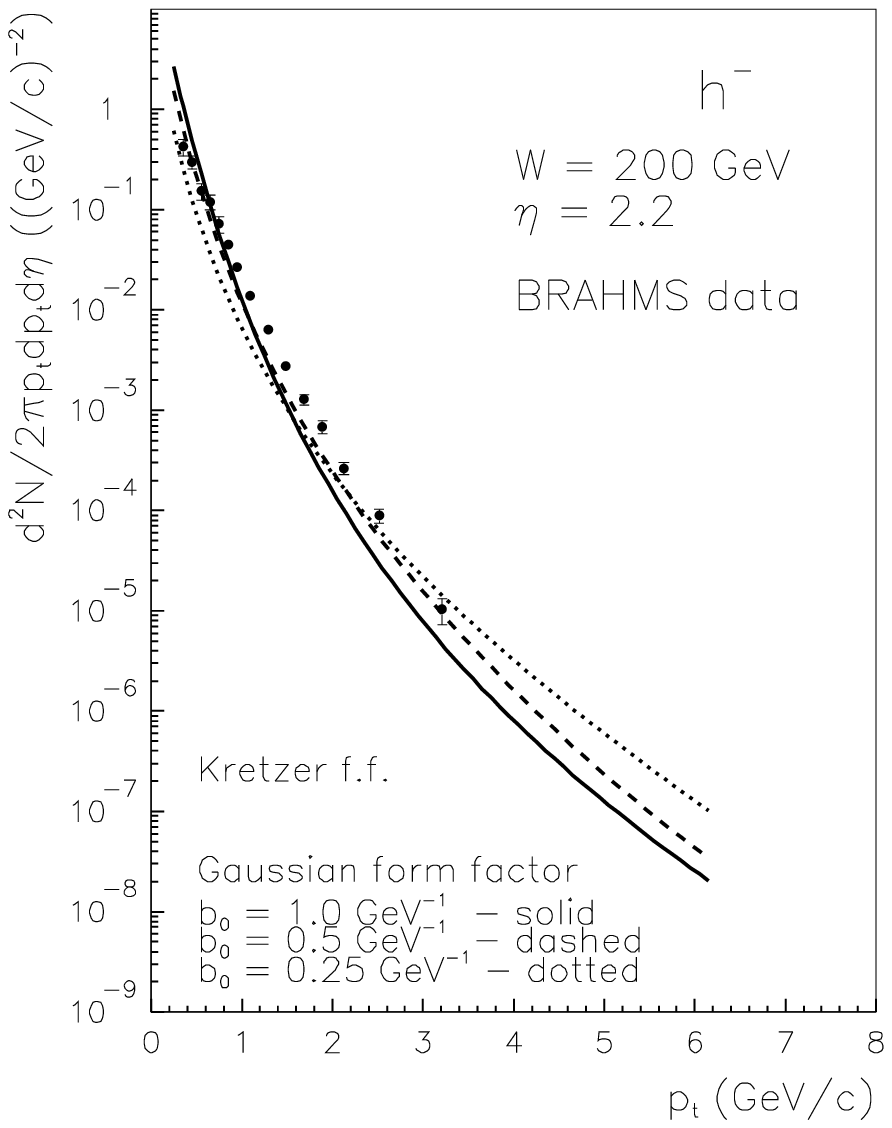}
\includegraphics[width=4.5cm]{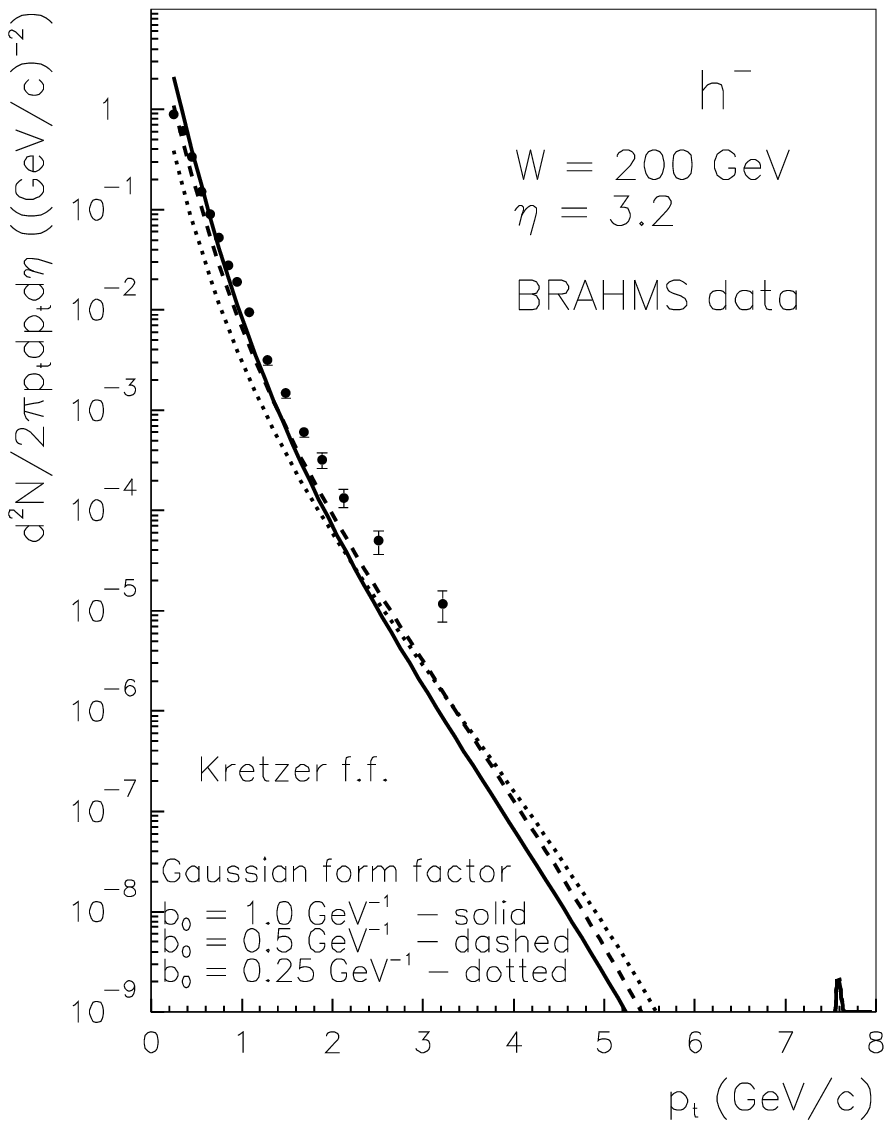}
\caption[*]{
Invariant cross section for charged particle production for
different pseudorapidities at W = 200 GeV.
The lines represent results obtained within our $k_t$-factorization
approach.
The BRAHMS collaboration experimental data \cite{BRAHMS_charged_hadrons}
are shown for comparison.
\label{fig:BRAHMS_our_b0}
}
\end{center}
\end{figure}

In Fig.\ref{fig:BRAHMS_our_b0} we present result of our calculations
at the RHIC energy W = 200 GeV again for different $b_0$
and Kretzer FFs \cite{Kretzer2000}.
While at large (pseudo)rapidities ($\eta$ = 2.2, 3.2) our
results (negative pions) are quite compatible with the BRAHMS data
for negative hadrons, there seems to be a missing strength at
more central (pseudo)rapidities ($\eta$ = 0.0, 1.0), i.e. in the case
when the sum of positively and negatively charged hadrons is measured.
It is not clear to us if the missing strength is due
to $p$ or $K^+$. 
Much more results and comparison with recent RHIC data will be shown
soon in \cite{CS05b}.

\section{CONCLUSIONS}

We have extended the approach of unintegrated gluon distributions
to particle production by including also unintegrated quark
and antiquark distributions.
We find that the new contributions are comparable to
the purely gluonic one at midrapidities
and dominate in the fragmentation region.
The new mechanisms are responsible for $\pi^+ - \pi^-$ asymmetry,
especially at SPS energies \cite{CS05,CS05b}.

Inclusive distributions in transverse momentum of pions for
both SPS and RHIC energies have been shown here for illustration.
A rather good agreement with experimental data is obtained without
introducing so-called K-factors and/or initial (internal) transverse
momenta by hand.
In our approach transverse momentum is generated to large extent
perturbatively by solving coupled equations for UPDF's. 
Therefore, in contrast to the standard collinear approach, in our
approach the range of applicability can be extended towards much
lower transverse momenta.

\end{document}